\documentclass[graphicx,aps,12pt]{article}

\input epsf

\newcommand{\be}{\begin{equation}}
\newcommand{\ee}{\end{equation}}
\newcommand{\bc}{\begin{center}}
\newcommand{\ec}{\end{center}}

\newcommand{\bea}{\begin{eqnarray}}
\newcommand{\eea}{\end{eqnarray}}

\def\lambdabar{\protect\@lambdabar}
\def\@lambdabar{%
\relax \bgroup
\def\@tempa{\hbox{\raise.73\ht0
\hbox to0pt{\kern.25\wd0\vrule width.5\wd0
height.1pt depth.1pt\hss}\box0}}%
\mathchoice{\setbox0\hbox{$\displaystyle\lambda$}\@tempa}%
{\setbox0\hbox{$\textstyle\lambda$}\@tempa}%
{\setbox0\hbox{$\scriptstyle\lambda$}\@tempa}%
{\setbox0\hbox{$\scriptscriptstyle\lambda$}\@tempa}%
\egroup }

\begin{document}

\thispagestyle{empty}
\renewcommand{\thefootnote}{\fnsymbol{footnote}}


\vspace{.8cm}

\begin{center}
{\bf\large PLASMA WAKEFIELD ACCELERATION FOR \\
ULTRA HIGH ENERGY COSMIC RAYS\footnote{Work supported by
Department of Energy contracts DE--AC03--76SF00515 and
DE-AC03-78SF00098.}}

\vspace{1cm} Pisin Chen\\ Stanford Linear
Accelerator Center \\ Stanford University, Stanford, CA 94309\\
\medskip
Toshiki Tajima\\ Advanced Photon Research Center \\ Japan Atomic
Energy
Research Institute, Kyoto, 619-0215 Japan\\
\medskip
Yoshiyuki Takahashi\\
Department of Physics, University of Alabama, Huntsville, AL 35899\\
\medskip
\end{center}
\vfill

\begin{center}
{\bf\large Abstract}
\end{center}

\begin{quote}
A cosmic acceleration mechanism is introduced which is based on
the wakefields excited by the Alfven shocks in a relativistically
flowing plasma, where the energy gain per distance of a test
particle is Lorentz invariant. We show that there exists a
threshold condition for transparency below which the accelerating
particle is collision-free and suffers little energy loss in the
plasma medium. The stochastic encounters of the random
accelerating-decelerating phases results in a power-law energy
spectrum: $f(\epsilon)\propto 1/\epsilon^2$. The environment
suitable for such plasma wakefield acceleration can be cosmically
abundant. As an example, we discuss the possible production of
super-GZK ultra high energy cosmic rays (UHECR) through this
mechanism in the atmosphere of gamma ray bursts. We show that the
acceleration gradient can be as high as $G\sim 10^{16}$eV/cm. The
estimated event rate in our model agrees with that from UHECR
observations.

\end{quote}

\vfill

\begin{center}
{\it Submitted to Phys. Rev. Lett.}  \\
\end{center}

\newpage
\pagestyle{plain}

Ultra high energy cosmic ray (UHECR) events exceeding the
Greisen-Zatsepin-Kuzmin (GZK) cutoff\cite{GZK} ($5\times
10^{19}$eV for protons originated from a distance larger than
$\sim 50$ Mps) have been found in recent
years\cite{FlysEye,AGASA,HiRes,Haverah}. Observations also
indicate a change of the power-law index in the UHECR spectrum
(events/area/time $\propto \epsilon^{-\alpha}$) from $\alpha\sim
3$ to a smaller value, at energy around $10^{18}-10^{19}$eV (the
so-called ``ankle"). These present an acute theoretical challenge
regarding their composition as well as their origin\cite{olinto}.

So far the theories that attempt to explain the UHECR can be
largely categorized into the ``top-down" and the ``bottom-up"
scenarios. In addition to relying on exotic particle physics
beyond the standard model, the main challenges of top-down
scenarios are their difficulty in compliance with the observed
event rates and the energy spectrum\cite{olinto}, and the
fine-tuning of particle lifetimes. The main challenges of the
bottom-up scenarios, on the other hand, are the GZK cutoff, as
well as the lack of an efficient acceleration
mechanism\cite{olinto}. To circumvent the GZK limit, several
authors propose\cite{weiler} that it is neutrinos, instead of
protons, that are the actual messenger. With the much weaker
(electroweak) interaction, neutrinos can propagate across a much
larger cosmic distance than the GZK radius. These ultra-high
energy neutrinos would eventually be annihilated by cosmic
background neutrinos and turn into Z-bosons. If this happens in
our surrounding cluster, the proton that is produced from the
Z-decay can reach the Earth. For such a cascade scenario to work,
it requires that the original particle, say protons, be several
orders of magnitude more energetic than the one eventually reaches
the Earth.

Even if the GZK-limit can be circumvented through the Z-burst,
the challenge for a viable acceleration mechanism remains, or
becomes even more acute. This is mainly because the existing
paradigm for cosmic acceleration, namely the Fermi
mechanism\cite{fermi}, as well as its variants, such as the
diffusive shock acceleration\cite{waxman}, are not effective in
reaching ultra high energies\cite{achterberg}. These acceleration
mechanisms rely on the random collisions of the high energy
particle against magnetic field domains or the shock media, which
necessarily induce increasingly more severe energy losses at
higher particle energies. There is therefore a ``diminishing
return" where the energy gain becomes evermore difficult, if not
impossible.

From the experience of terrestrial particle accelerators, we learn
that it takes several qualifications for an accelerator to operate
effectively. First, the particle should gain energy through the
interaction with the longitudinal electric field of a subluminous
($v\leq c$) electromagnetic (EM) wave. In such a setting the
accelerated particle can gain energy from the field over a
macroscopic distance, much like how a surfer gains momentum from
an ocean wave. It is important to note that such a longitudinal
field is Lorentz invariant, meaning that the acceleration gradient
is independent of the instantaneous energy of the accelerating
particle. Second, such a particle-field interaction should be a
non-collisional process. This would help to avoid severe energy
loss through inelastic scatterings. Third, to avoid excessive
synchrotron radiation loss, which scales as particle energy
squared, the accelerating particle should avoid any drastic
bending beyond certain energy regime. We believe that these
qualifications for terrestrial accelerators are also applicable to
celestial ones.

Although they are still in the experimental stage, the ``plasma
wakefield accelerator" concepts\cite{tajima-dawson,chen85},
promise to provide all the conditions stated above. Plasmas are
capable of supporting large amplitude electro-static waves with
phase velocities near the speed of light. Such collective waves,
or ``wakefields", can be excited by highly concentrated,
relativistic EM energies such as lasers\cite{tajima-dawson} and
particle beams\cite{chen85}. A trailing particle can then gain
energy by riding on this wakefield. Although hard scatterings
between the accelerating particle and the plasma medium is
inevitable, under appropriate conditions, as we will demonstrate
below, the particle can be collision-free.

In this Letter we demonstrate that magneto-shocks (Alfven shocks)
in a relativistic plasma flow can also excite large amplitude
plasma wakefields, which in turn can be highly efficient in
accelerating ultra high energy particles. We note that the
physical conditions required for such acceleration are cosmically
abundant and therefore generic. When applying this mechanism to
the atmosphere of gamma ray bursts (GRBs), we show that protons
can be accelerated to energies much beyond ZeV ($10^{21}$ eV) with
a power-law spectrum.

 It is well-known that an ordinary Alfven wave propagating in a
stationary magnetized plasma has a velocity
\begin{equation}
v_{_A}=\frac{eB_0}{\sqrt{4\pi m_i n_p}}\ ,
\end{equation}
which is typically much less than the speed of light. Here $B_0$
is the longitudinal magnetic field and $n_p$ is the density of the
magnetized plasma. Such a slow wave is ordinarily not suitable for
accelerating relativistic particles. The situation changes when
the plasma as a whole moves with a relativistic bulk velocity
$V_p\leq c$. The standard method of obtaining the linear
dispersion relation of waves in a magnetized plasma leads to
\begin{equation}
\frac{k_z^2c^2}{\omega^2}=1-\frac{1}{\Gamma_p}\frac{(\omega_{pi}^2+
\omega_{pe}^2)(1-V_pk/\omega)}{(\omega-V_pk\pm
\omega_{Bi}/\Gamma_p)(\omega-V_pk\mp \omega_{Be}/\Gamma_p)}\ ,
\end{equation}
where $k$ and $\omega$ are the wave number and the frequency of
the EM wave, respectively, $\omega_{pi,pe}=(4\pi
e^2n_p/m_{i,e})^{1/2}$ are the plasma frequencies for ions and
electrons, and $\omega_{Bi,Be}=(eB_0/m_{i,e})^{1/2}$ are the ion
and electron cyclotron frequencies. Here $\Gamma_p$ is the Lorentz
factor of the bulk plasma flow. Figure 1 shows the dispersion
relations of various transverse EM waves that propagate along the
direction of $B_0$ with and without the plasma bulk flow $V_p$. In
Fig. 1(a) we see that outside the lightcone (superluminous, or
$v_{ph}> c$) lie the regular EM waves, whose asymptotic dispersion
is $\omega=kc$. Within the lightcone (subluminous), there are two
additional branches, the whistler wave (an electron branch mode)
and the Alfven wave whose frequency remains quite low and its
electric field is much smaller than the magnetic one, i.e.,
$E/B=v_{_A}/c\ll 1$ in the absence of flow. In the case where the
bulk flow of the plasma approaches the speed of light, however,
both the whistler and Alfven waves now acquire a phase velocity
close to $c$ from below (see Fig. 1(b)) and enhances the ratio of
$E/B$ to $\sim V_p/c \leq 1$. It is important to emphasize that as
the Alfven wave dispersion approaches the lightcone, it becomes
indistinguishable from a bona fide EM wave, and is therefore able
to excite relativistic plasma wakefields, much like lasers and
particle beams. Preliminary results from simulations indicate that
such relativistic Alfven waves can indeed excite plasma
wakefields\cite{romenesko}. It is also important to realize that
in this relativistic flow the excited wakefields are all in one
direction, which contributes to the unidirectional acceleration.
With our applications to astrophysical problems in mind, the
Alfven-wave-plasma interaction relevant to us is in the nonlinear
regime.

The plasma wakefield in the nonlinear regime has been
well-studied\cite{esarey}. The nonlinearity is determined by the
driving EM wave's {\it ponderomotive} potential, which is governed
by its normalized vector potential $a_0=eE/mc\omega$. When this
parameter exceeds unity, nonlinearity is
strong\cite{tajima-dawson} so that additional important physics
incurs. For a stationary plasma, the maximum field amplitude that
the plasma wakefield can support is
\begin{equation}
E_{\rm max}\approx E_{\rm wb}a_0=\frac{m_ec\omega_p}{e}a_0\ ,
\end{equation}
which is enhanced by a factor $a_0$ from the cold wavebreaking
limit (the naively assumed maximum field), $E_{\rm
wb}=m_ec\omega_p/e$, of the linear regime. In a relativistic
plasma flow with a Lorentz factor $\Gamma_p$, the cold
wavebreaking field is reduced by a factor $\Gamma_p^{1/2}$ due to
Lorentz contraction. The maximum ``acceleration gradient" $G$
experienced by a singly-charge particle riding on this plasma
wakefield is then
\begin{equation}
G=eE'_{\rm max}\approx a_0m_ec^2\sqrt{\frac{4\pi r_e
n_p}{\Gamma_p}}\ .
\end{equation}
The plasma wavelength, in the mean time, is stretched also by a
factor $a_0$ from that in the linear regime. So in a plasma flow
the wavelength is
\begin{equation}
\lambda_{pN}=\frac{2}{\pi}a_0\lambda'_p\approx
a_0\sqrt\frac{\pi\Gamma_p}{r_en_p}\ ,
\end{equation}
where $r_e=e^2/m_ec^2=2.8\times 10^{-13}$cm is the classical
electron radius.

To determine the maximum possible energy gain, we need to know how
far can a test particle be accelerated. At ultra high energies
once the test particle encounters a hard scattering or bending,
the hard-earned kinetic energy would most likely be lost. The
scattering of an ultra high energy proton with the background
plasma is dominated by the proton-proton collision. Existing
laboratory measurements of the total $pp$ cross section scales
roughly as $\sigma_{pp}=\sigma_0\cdot\{1+6.30\times
10^{-3}[\log(s)]^{2.1}\}$, where $\sigma_0\approx 32$mb and the
center-of-mass energy-squared, $s$, is given in (GeV)$^2$. In our
system, even though the UHE protons are in the ZeV regime, the
center-of-mass energy of such a proton colliding with a comoving
background plasma proton is in the TeV range, so it is safe to
ignore the logarithmic dependence and assume a constant total
cross section, $\sigma_{pp}\sim \sigma_0\sim 30$ mb in the ZeV
energy regime. Since in astrophysical settings an out-bursting
relativistic plasma dilutes as it expands radially, its density
scales as $n_p(r)=n_{p0}(R_0/r)^2$, where $n_{p0}$ is the plasma
density at a reference radius $R_0$ . The proton mean-free-path
can be determined by integrating the collision probability up to
unity,
\begin{equation}
1 =\int_{R_0}^{R_0+L_{\rm
mfp}}\frac{\sigma_{pp}n_p(r)}{\Gamma_p}dr=\int_{R_0}^{R_0+L_{\rm
mfp}}\frac{\sigma_{pp}n_{p0}}{\Gamma_p}\frac{R_0^2}{r^2}dr\ .
\end{equation}
We find
\begin{equation}
1 =
\frac{\sigma_{pp}n_{p0}R_0}{\Gamma_p}\Big[1-\frac{R_0}{R_0+L_{\rm
mfp}}\Big]\ .
\end{equation}
Since $L_{\rm mfp}$ is positive definite, $0<[1-R_0/(R_0+L_{\rm
mfp})]< 1$. Therefore the solution to $L_{\rm mfp}$ does not exist
unless the coefficient, $\sigma_{pp}n_{p0}R_0/\Gamma_p>1$. That is
there exists a threshold condition below which the system is
collision-free:
\begin{equation}
\frac{\sigma_{pp}n_{p0}R_0}{\Gamma_p}=1\ .
\end{equation}
When a system is below this threshold, a test particle can in
principle be accelerated unbound. In practice, of course, other
secondary physical effects would eventually intervene.

In a terrestrial accelerator, the wakefields are coherently
excited by the driving beam, and the accelerating particle would
ride on the same wave crest over a macroscopic distance. There the
aim is to produce near-monoenergetic final energies (and tight
phase-space) for high energy physics and other applications. In
astrophysical settings, however, the drivers, such as the Alfven
shocks, will not be so organized. A test particle would then face
random encounters of accelerating and decelerating phases of the
plasma wakefields excited by Alfven shocks.

The stochastic process of the random acceleration-deceleration can
be described by the distribution function $f(\epsilon,t)$ governed
by the Chapman-Kolmogorov equation\cite{mima,yoshi}
\begin{equation}
\frac{\partial}{\partial t}f
=\int_{-\infty}^{+\infty}d(\Delta\epsilon)
W(\epsilon-\Delta\epsilon,\Delta\epsilon)f(\epsilon-\Delta\epsilon,t)
-\int_{-\infty}^{+\infty}d(\Delta\epsilon)
W(\epsilon,\Delta\epsilon)f(\epsilon,t)-\nu(\epsilon)f(\epsilon,t)\ .
\end{equation}
The first term governs the probability per unit time of a particle
``sinking" into energy $\epsilon$ from an initial energy
$\epsilon-\Delta\epsilon$ while the second term that ``leaking"
out from $\epsilon$. The last term governs the dissipation due to
collision or radiation, or both. As we will demonstrate later, the
astrophysical environment that we invoke for the production of
UHECR is below the collision threshold condition, and so
accelerating particles are essentially collision-free.

The radiation loss in our system is also negligible. As discussed
earlier, in a relativistic flow the transverse $E$ and $B$ fields
associated with the Alfven shock are near equal in magnitude.
Analogous to that in an ordinary EM wave, an ultra relativistic
particle (with a Lorentz factor $\gamma$) co-moving with such a
wave will experience a much suppressed bending field, by a factor
$1/\gamma^2$. Furthermore, the plasma wakefield acceleration takes
place in the region that trails behind the shock (and not in the
bulk of the shock) where the accelerating particle in effect sees
only the longitudinal {\it electrostatic} field colinear to the
particle motion\cite{esarey}. We are therefore safe to ignore the
radiation loss entirely as well. We can thus ignore the
dissipation term in the Chapman-Komogorov equation and focus only
on the purely random plasma wakefield acceleration-deceleration.

Assuming that the energy gain per phase encounter is much less
than the final energy, i.e., $\Delta\epsilon \ll \epsilon$, we
Taylor-expand
$W(\epsilon-\Delta\epsilon,\Delta\epsilon)f(\epsilon-\Delta\epsilon)$
around  $W(\epsilon,\Delta\epsilon)f(\epsilon)$ in the sink term
and reduce Eq.(9) to the Fokker-Planck equation
\begin{equation}
\frac{\partial}{\partial t}f =\frac{\partial}{\partial \epsilon}
\int_{-\infty}^{+\infty}d(\Delta\epsilon)\Delta\epsilon
W(\epsilon,\Delta\epsilon)f(\epsilon,t)
+\frac{\partial^2}{\partial
\epsilon^2}\int_{-\infty}^{+\infty}d(\Delta\epsilon)\frac
{\Delta\epsilon^2}{2}W(\epsilon,\Delta\epsilon)f(\epsilon,t)\ .
\end{equation}

We now assume the following properties of the transition rate
$W(\epsilon,\Delta\epsilon)$ for a purely stochastic process:

a) $W$ is an even function;

b) $W$ is independent of $\epsilon$;

c) $W$ is independent of $\Delta\epsilon$.

Property a) follows from the fact that in a plasma wave there is
an equal probability of gaining and losing energy. In addition,
since the wakefield amplitude is Lorentz invariant, the chance of
gaining a given amount of energy, $\Delta\epsilon$, is independent
of the particle energy $\epsilon$. Finally, under a purely
stochastic white noise, the chance of gaining or losing any amount
of energy is the same. Based on these arguments we deduce that
\begin{equation}
W(\epsilon,\Delta\epsilon)=\frac{1}{2c\tau^2G} \ ,
\end{equation}
where $\tau$ is the typical time of interaction between the test
particle and the random waves and $G$ is the maximum acceleration
gradient (cf. Eq.(4)). We note that there is a stark departure of
the functional dependence of $W$ in our theory from that in
Fermi's mechanism, in which the energy gain $\Delta\epsilon$ per
encounter scales linearly and quadratically in $\epsilon$ for the
first-order and second-order Fermi mechanism, respectively.

To look for a stationary distribution, we put $\partial
f/\partial t=0$. Since $W$ is an even function, the first term on
the RHS in Eq.(10) vanishes. To ensure the positivity of particle
energies before and after each encounter, the integration limits
are reduced from $(-\infty,+\infty)$ to $[-\epsilon,+\epsilon]$,
and we have
\begin{equation}
\frac{\partial^2}{\partial
\epsilon^2}\int_{-\epsilon}^{+\epsilon}d(\Delta\epsilon)\frac
{\Delta\epsilon^2}{2}W(\epsilon,\Delta\epsilon)f(\epsilon)=0\ .
\end{equation}
Inserting $W$ from Eq.(11), we arrive at the energy distribution
function that follows power-law scaling,
\begin{equation}
 f(\epsilon)=\frac{\epsilon_0}{\epsilon^2}\ ,
\end{equation}
where the normalization factor $\epsilon_0$ is taken to be the
mean energy of the background plasma proton, $\epsilon_0\sim
\Gamma_p m_pc^2$. The actually observed UHECR spectrum is expected
to be degraded somewhat from the above idealized, theoretical
power-law index, $\alpha=2$, not only due to possible departure of
the reality from the idealized model, but also due to additional
intermediate cascade processes that transcend the original UHE
protons to the observed UHECRs.

We note that a power-law energy spectrum is generic to all purely
stochastic, collisionless acceleration processes. This is why both
the first and the second order Fermi mechanisms also predict
power-law spectrum, if the energy losses, e.g., through inelastic
scattering and radiation (which are severe at ultra high
energies), are ignored. The difference is that in the Fermi
mechanism the stochasticity is due to random collisions of the
test particle against magnetic walls or the shock medium, which
necessarily induce {\it reorientation} of the momentum vector of
the test particle after every diffusive encounter, and therefore
should trigger inevitable radiation loss at high energies. The
stochasticity in our mechanism is due instead to the random
encounters of the test particle with different
accelerating-decelerating {\it phases}. As we mentioned earlier,
the phase vector of the wakefields created by the Alfven shocks in
the relativistic flow is nearly unidirectional. The particle's
momentum vector, therefore, never changes its direction but only
magnitude, and is therefore radiation free in the energy regime
that we consider for proton acceleration.

We now apply our acceleration mechanism to the problem of UHECR.
GRBs are by far the most violent release of energy in the
universe, second only to the big bang itself. Within seconds (for
short bursts) about $\epsilon_{\rm GRB}\sim 10^{52}{\rm erg}$ of
energy is released through gamma rays with a spectrum that peaks
around several hundred keV. Existing models for GRB, such as the
relativistic fireball model\cite{meszaros}, typically assume
neutron-star-neutron-star (NS-NS) coalescence as the progenitor.
Neutron stars are known to be compact ($R_{\rm NS}\sim O(10){\rm
km}$) and carrying intense surface magnetic fields ($B_{\rm
NS}\sim 10^{12}{\rm G}$). Several generic properties are assumed
when such compact objects collide. First, the collision creates
sequence of strong magneto-shocks (Alfven shocks). Second, the
tremendous release of energy creates a highly relativistic
out-bursting fireball, most likely in the form of a plasma.

The fact that the GRB prompt (photon) signals arrive within a
brief time-window implies that there must exists a threshold
condition in the GRB atmosphere where the plasma becomes optically
transparent beyond some radius $R_0$ from the NS-NS epicenter.
Applying Eq.(8) to the case of out-bursting GRB photons, this
condition means
\begin{equation}
\frac{\sigma_c n_{p0}R_0}{\Gamma_p}= 1\ ,
\end{equation}
where $\sigma_c=(\pi r_e^2)(m_e/\omega_{\rm GRB})
[\log(2\omega_{\rm GRB}/m_e)+1/2]\approx 2\times 10^{-25}{\rm
cm}^2$ is the Compton scattering cross section. Since
$\sigma_{pp} < \sigma_c$, the UHECRs are also collision-free in
the same environment. There is clearly a large parameter space
where this condition is satisfied. To narrow down our further
discussion, it is not unreasonable to assume that $R_0\sim
O(10^4){\rm km}$. A set of self-consistent parameters can then be
chosen: $n_{p0}\sim 10^{20}{\rm cm}^{-3}, \Gamma_p\sim 10^4$, and
$\epsilon_0\sim 10^{13}{\rm eV}\equiv \epsilon_{13}$.

To estimate the plasma wakefield acceleration gradient, we first
derive the value for the $a_0$ parameter. We believe that the
megneto-shocks constitute a substantial fraction, say $\eta_a\sim
10^{-2}$, of the total energy released from the GRB progenitor.
The energy Alfven shocks carry is therefore $\epsilon_A\sim
10^{50}{\rm erg}$. Due to the pressure gradient along the radial
direction, the magnetic fields in Alfven shocks that propagate
outward from the epicenter will develop sharp discontinuities and
be compactified\cite{Jeffrey-Taniuti}. The estimated shock
thickness is $\sim O(1){\rm m}$ at $R_0\sim O(10^4){\rm km}$. From
this and $\epsilon_A$ one can deduce the magnetic field strength
in the Alfven shocks at $R_0$, which gives $B_A\sim 10^{10}{\rm
G}$. This leads to $a_0=eE_A/mc\omega_A\sim 10^9$. Under these
assumptions, the acceleration gradient $G$ ({\it cf.} Eq.(4)) is
as large as
\begin{equation}
G\sim a_0 mc^2\sqrt{\frac{4\pi r_e}{\sigma_c R_0}}\sim 10^{16}
\Big(\frac{a_0}{10^9}\Big)\Big(\frac{10^9{\rm cm}
}{R_0}\Big)^{1/2}{\rm eV/cm}\ .
\end{equation}

Although the UHE protons can in principle be accelerated unbound
in our system, the ultimate maximum reachable energy is determined
by the conservation of energy and our assumption on the population
of UHE protons. Since it is known that the coupling between the
ponderomotive potential of the EM wave and the plasma wakefield is
efficient, we assume that the Alfven shock energy is entirely
loaded to the plasma wakefields after propagating through the
plasma. Furthermore, we assume that the energy in the plasma
wakefield is entirely reloaded to the UHE protons through the
stochastic process. Thus the highest possible UHE proton energy
can be determined by energy conservation
\begin{equation}
\epsilon_{\rm UHE}\sim \epsilon_{\rm A}\sim \eta_a\epsilon_{\rm
GRB}\sim N_{\rm UHE} \int_{\epsilon_{13}}^{\epsilon_m}\epsilon
f(\epsilon)d\epsilon \ .
\end{equation}
which gives
\begin{equation}
\epsilon_m=\epsilon_{13} \exp({\eta_a\epsilon_{\rm GRB}/N_{\rm
UHE}\epsilon_{13}}) \ .
\end{equation}
This provides a relationship between the maximum possible energy,
$\epsilon_m$, and the UHE proton population, $N_{\rm UHE}$. We
assume that $\eta_b\sim 10^{-2}$ of the GRB energy is consumed to
create the bulk plasma flow, i.e., $\eta_b\epsilon_{\rm GRB}\sim
N_p\Gamma_pm_pc^2\sim N_p\epsilon_{13}$, where $N_p$ is the total
number of plasma protons. We further assume that $\eta_c\sim
10^{-2}$ of the plasma protons are trapped and accelerated to UHE,
i.e., $N_{\rm UHE}\sim \eta_cN_p$. Then we find $\epsilon_m\sim
\epsilon_{13} \exp(\eta_a/\eta_b\eta_c)$. We note that this
estimate of $\epsilon_m$ is exponentially sensitive to the ratio
of several efficiencies, and therefore should be handled with
caution. If the values are indeed as we have assumed,
$\eta_a/\eta_b\eta_c\sim O(10^2)$, then $\epsilon_m$ is
effectively unbound until additional limiting physics enters.
Whereas if the ratio is $\sim O(10)$ instead, the UHE cannot even
reach the ZeV regime. The validity of our assumed GRB efficiencies
then relies on the consistency check against observations.

In addition to the energy production issue, equally important to a
viable UHECR model is the theoretical estimate of the UHECR event
rates. The NS-NS coalescence rate is believed to be about 10
events per day in the entire Universe\cite{Piran(GRB),LISA}. This
frequency is consistent with the observed GRB events, which is on
the order of $f_{\rm GRB}\sim 10^{3.5}$ per year.

In the Z-burst scenario an initial neutrino energy above
$10^{21}{\rm eV}$\cite{weiler} or $10^{23} {\rm eV}$\cite{gelmini}
is required (depending on the assumption of the neutrino mass) to
reach the Z-boson threshold. For the sake of discussion, we shall
take the necessary neutrino energy as $\epsilon_{\nu}> 10^{22}{\rm
eV}$. Such ultra high energy neutrinos can in principle be
produced through the collisions of UHE protons with the GRB
background protons: $pp\to \pi+X\to \mu+\nu+X$. All UHE protons
with energy $\epsilon_{>22}\geq 10^{22}{\rm eV}$ should be able
to produce such neutrinos. The mean energy (by integrating over
the distribution function $f(\epsilon)$) of these protons is
$\langle \epsilon_{>22}\rangle \sim O(100)\epsilon_{22}$.
Therefore the multiplicity of neutrinos per UHE proton is around
$\mu_{(p\to\nu)}\sim O(10)-O(100)$. At the opposite end of the
cosmic process, we also expect multiple hadrons produced in a
Z-burst. The average number of protons that Z-boson produces is
$\sim 2.7$\cite{who}. Finally, the population of UHE protons above
$10^{22} {\rm eV}$ is related to the total UHE population by
$N_{>22}\sim (\epsilon_{13}/\epsilon_{22})N_{\rm UHE}\sim
\eta_b\eta_c\epsilon_{\rm GRB}/\epsilon_{22}$.

Putting the above arguments together, we arrive at our theoretical
estimate of the expected UHECR event rate on earth,
\begin{eqnarray}
N_{\rm UHECR}(>10^{20}{\rm eV})&=& f_{\rm
GRB}\mu_{(p\to\nu)}\mu_{(Z \to p)}N_{>22}\frac{1}{4\pi R_{\rm GRB}^2} \nonumber \\
 &\sim & f_{\rm GRB}
\mu_{(p\to\nu)} \mu_{(Z \to p)} \eta_b\eta_c\frac{\epsilon_{\rm
GRB}} {\epsilon_{22}} \frac{1}{4\pi R_{\rm GRB}^2} \ .
\end{eqnarray}
The typical observed GRB events is at a redshift $z\sim O(1)$, or a
distance $R_{\rm GRB}\sim 10^{22}$km. Our estimate of observable
UHECR event rate is therefore
\begin{equation}
N_{\rm UHECR}(>10^{20}{\rm eV})= O(1)/100{\rm km}^2/{\rm yr}/{\rm
sr}\ ,
\end{equation}
which is consistent with observations, or in turn this observed
event rate can serve as a constraint on the various assumptions
of our specific GRB model.

We have demonstrated that plasma wakefields excited by Alfven
shocks in a relativistic plasma flow can be a very efficient
mechanism for cosmic acceleration, with a power-law energy
spectrum. When invoking GRBs as the sites for UHECR production
with a set of reasonable assumptions, we show that our estimated
UHECR event rate is consistent with observations. This cosmic
acceleration mechanism is generic, and can in principle be applied
to other astrophysical phenomena, such as {\it
blazars}\cite{blazar}. It is generally believed that the AGN jets
are relativistic plasmas. The observed ``lumps", or density
concentrations, in the jet may well serve as the driver to excite
plasma wakefields. These wakefields can accelerate electrons as
well as protons to multi-TeV energies. Bent by the confining
helical magnetic fields in the jet, these high energy electrons
can radiate hard photons in the TeV range, while the protons can
cascade into high energy neutrinos. We will present a more
detailed discussion on blazars in a separate paper.

We appreciate helpful discussions with J. Arons, R. Blandford, P.
Meszaros.

\newpage
\begin{figure}[htb]
\begin{center}
\includegraphics{8617A9a.eps}
\end{center}
\vspace{6in} \caption{The dispersion relations for stationary and
relativistic plasma flows.} \label{fig:1}
\end{figure}

\end{document}